# Tecnologías móviles en la Educación

# Mobile Technologies in Education

Serna Poot Daniel
*Universidad Veracruzana – Universidad Da Vinci*
dserna@udavinci.edu.mx, dserna@uv.mx

**Resumen**
El crecimiento de usuarios de *smartphones* a nivel global es un factor que los tecnólogos educativos no deben ignorar. Este mercado en constante crecimiento desembocará eventualmente en un aprendizaje ubicuo (u-learning). El desarrollo de contenidos específicos deberá ser reemplazado por el diseño de contenidos a lenguajes compatibles con tecnologías escalables y que lleguen de manera atractiva a las manos de los estudiantes.

Palabras Clave: Educación y Tecnología, Ciencia, Tecnología y sociedad

Abstract

The growth of smartphone users globally is a factor that educational technologists should not ignore. This growing market will eventually lead to ubiquitous learning (u-learning). The development of specific content must be replaced by content design scalable technologies compatible languages and arrive attractively into the hands of students.

**Introducción**

El avance de las herramientas tecnológicas ha sido una constante en el desarrollo humano desde que se tiene conciencia colectiva. Las innovaciones han ido acelerándose como si fuera la caída de un objeto en un vacío que parece no tener fin y la lista de nuevos dispositivos crece a un ritmo exponencial.

Estas innovaciones tecnológicas han ido de la mano de la ciencia, la cual se nutre de la curiosidad humana. En esta curiosidad de saber y conocer de las cosas que nos rodean, se han ido descubriendo distintos materiales y componentes con diversas propiedades físicas y eléctricas. También se ha profundizado en el desarrollo de teorías que han llevado a nuevas aplicaciones de adelantos ya conocidos. Una de esas innovaciones tecnológicas sobresalientes es el teléfono

**Reseña del desarrollo de la telefonía**

El invento del teléfono se produjo en 1857 de la mente del italiano Antonio Meucci, y no de Alexander Graham Bell, como se cree generalmente (Carrol, 2002). Sobre este caso hay mucho que escribir, pero por el momento basta con mencionar que Bell tenía acceso a los archivos y muestras del invento desde 1870, cuando trabajaba en la Western Union Telegraph Company, compañía que recibió la muestra de manos de Meuchi. Bell logró juntar los $250 dólares que se necesitaban para patentar el invento antes que lo pudiera hacer Meucci (Carrol, 2002), por lo que a partir de ese momento inició una batalla legal que terminó en 2002 mucho después de la muerte de estos dos protagonistas (Carrol, 2002).

El "telégrafo parlante", como lo llamó Meucci, tenía como finalidad ayudar a su esposa, quien padecía de reumatismo y no podía levantarse, por lo que el inventor creo este artilugio para que ella lo pudiera llamar desde su cuarto que se encontraba en un segundo piso hasta su oficina (Lozano, 2002).

El invento inicialmente consistía en un par de diafragmas embobinadas, hilo de cobre y una pequeña corriente directa de 24 voltios. Hoy en día, el invento técnicamente sigue siendo igual, aunque se han mejorado los diafragmas, el embobinado es ahora más fino y la corriente que se requiere es ahora menor a los 6v, estando descolgado el auricular (Ortiz Zárate, 2010).

Las innovaciones más sobresalientes en la historia de la telefonía tienen que ver con la mejora de la calidad de la trasmisión y recepción de la voz humana, pero también con su funcionalidad: el uso de micrófonos de carbono y posteriormente de condensador, marcación por pulsos y más adelante por tonos.

Al aumentar el número de usuarios, las compañías telefónicas se dieron cuenta de que trasmitir tanta información por solamente dos hilos de cobre resultaba insuficiente, se dieron a la tarea de crear protocolos de comunicación con los cuales se podía trasmitir mucha más información.

Aun con estos nuevos protocolos, se dieron cuenta que se tenían que usar nuevos medios de transporte, encontrándolos en los hilos de fibra óptica y luz láser, con la cual es posible trasmitir no solamente voz, sino canales de datos en los que pueden trasmitirse video y audio, mediante hardware especializado que modula y demodula (modem) la información mediante computadoras para que finalmente llegue la señal de forma nuevamente audible.

Dentro de esos avances tecnológicos, la telefonía celular ocupa un lugar importante en nuestros días. Esta tecnología ha tenido algunas variantes muy importantes, como lo son los casos de la telefonía celular, las videoconferencias, los mensajes de texto sobre líneas telefónicas, la telefonía IP y el *streaming* de video por IP.

**La telefonía celular**

El caso de la telefonía celular merece atención especial, puesto que el teléfono celular de hoy en día es la suma de diferentes tecnologías y patentes, que hacen posible las diferentes capacidades que cada aparato consigue.

Este invento nació como muchos otros aparatos, de la segunda Guerra Mundial. En ese entonces (1940), la compañía Motorola creó un aparato llamado *Handie Talkie* para comunicar a las tropas a distancias relativamente cortas (US Army, 1955). La comunicación que se lograba era de punto a punto, mediante ondas de radio que no superaban los 60 Hz de frecuencia. Poco después, en 1942 Hedy Lamarr, ideó y patentó un sistema de conmutación de frecuencias, que daría nacimiento a las tecnologías WiFi, PCM y 3G (Walfer, 2013).

En 1983, la compañía Motorola creó el teléfono DynaTac 8000X, el cual fue el primer teléfono móvil del mundo. Pesaba 800 gramos, costaba casi $4 mil dólares y su batería no duraba más de una hora. Pero se vendieron más de 300 mil unidades en un sólo año, por lo que se superaron ampliamente las expectativas que la compañía tenía sobre su venta (Velasco, 2012).

Este tipo de telefonía se le denominó "celular", porque se basó en células de trasmisión: antenas posicionadas interconectadas en red que brindaban servicio a los teléfonos móviles para hablar entre los usuarios, mismos que pagaban altas cuotas por contar con comunicación móvil basada también en ondas de radio, ahora en frecuencias más altas que las de su predecesor.

A raíz de este gran avance, diversas compañías se dedicaron a fabricar nuevos y mejores teléfonos celulares, con cada vez más funciones y propósitos.

Además del desarrollo del hardware, mucho tiempo y dinero se ha invertido para desarrollar el software que controla el hardware que interviene el flujo de comunicaciones. Al surgir la telefonía IP o telefonía a través de internet, muchos programadores decidieron que era posible crear plataformas telefónicas basadas en software libre, como por ejemplo Asterix, que está basado en el sistema operativo Linux (Meggelen, 2006).

La filosofía de compartir los conocimientos para que pasen a ser propiedad de la humanidad (Stallman, 2008) impidiendo la privatización de los conocimientos y poniéndolos al alcance de quien lo desee, permite usar nuestro hardware sin la necesidad de adquirir un costoso software privativo.

Gracias al software libre y al hardware cada vez más novedoso, es posible contar con teléfonos celulares cada vez más "inteligentes" o "smartphone" con mayores posibilidades en cada nueva edición.

El primer smartphone fue el "Simon", de IBM y "…fue el primer intento real de la industria tecnológica de crear un teléfono con algo más…" (Jan, 2011, pág. 2), contando con servicios de voz y datos, por lo que el equipo funcionaba como un teléfono móvil, asistente personal con calendario, libreta de direcciones, reloj mundial, calculadora, bloc de notas, correo electrónico y juegos. Podía servir además como una máquina de fax y contaba ya con pantalla táctil.

Después llegaron las PDAs (o Asistentes Personales Digitales, por sus siglas en inglés) de la compañía Palm Pilot, mismos que acostumbraron a la gente a llevar consigo sus datos personales, aunque inicialmente no eran teléfonos celulares, más adelante su sistema operativo sirvió para el desarrollo de teléfonos inteligentes.

En 1998, Nokia desarrolló un dispositivo llamado 9110 Comunicator, el cual contaba con pantalla táctil y teclado físico QWERTY. No era muy poderoso, pero su diseño fue copiado por otros dispositivos, sentando las bases para el uso de teclados tipo computadora. Nokia también desarrolló un sistema operativo para sus teléfonos llamado Symbian.

A fines de los 90s, la compañía canadiense *Research in Motion*, que distribuía localizadores personales, lanzó el Blackberry 5810, su primer dispositivo con el que se podía recibir correos electrónicos y permitía navegar en internet. Este dispositivo no tenía altavoces, pero permitía llamadas usando auriculares (Jan, 2011).

En 2003, se popularizó el uso de los dispositivos Palm Treo, los cuales ya contaban con 32 megabytes de RAM, admitían almacenamiento extraíble, se permitía la instalación de aplicaciones desarrollados por terceros, además de contar con agendas, recordatorios, edición de documentos, navegación por internet, acceso a redes WiFi, pantalla táctil y reproducción de archivos multimedia, así como cámara fotográfica, permitiendo además la grabación de video.

Más adelante, la compañía Palm desarrolló versiones del modelo Treo con Windows Mobile, un sistema operativo desarrollado por Microsoft y que no llegó a distribuirse ampliamente. Posteriormente (2010) la compañía Hewlett Packard adquiere a la compañía Palm, desmembrándola y desapareciendo la demanda por el producto en un tiempo considerablemente corto.

Fue hasta el 2007, en el que la compañía Apple, que había triunfado con su dispositivo reproductor de multimedia iPod, que se aventuró en el mundo de la telefonía celular con el iPhone. Algunas de las grandes innovaciones fueron en el hardware fueron su capacidad de memoria, su pantalla de cristal de alta definición, la reducción de su teclado a un solo botón, GPS (Geo Posicionador Satelital) y sus dos cámaras de video al frente y atrás. Las innovaciones del software fueron tal vez más importantes: por primera vez se usa en un celular una variante de un sistema operativo diseñado para una computadora Mac OSX. Los iPhone tienen un kernel basado en Unix, un sistema operativo muy conocido por su robustez y su estabilidad, pero sobre todo por su seguridad. Otra de las innovaciones que incluyó fue la posibilidad de adquirir aplicaciones en una tienda virtual vinculada al dispositivo. Por todas estas innovaciones, la revista Time declaró al iPhone el invento del año (Grossman, 2007).

En 2007, la compañía Google lanzó a la venta un Smartphone basado en un sistema operativo propio al que le llamó Android. Este sistema operativo está basado en Linux, un clon de Unix, por lo que los dispositivos Android y los iPhone pueden ser considerados "primos hermanos", aunque el iPhone se distingue por su estabilidad y velocidad en la ejecución de aplicaciones. La principal característica de Android es que puede ser instalado en numerosos dispositivos y de distintas marcas.

A pesar de haber tenido un lanzamiento relativamente "flojo", Android después de cinco años ha obtenido una importante cuota del mercado, llegando a superar en número de aplicaciones al AppleStore en 2013 (Ver Ilustración 1).

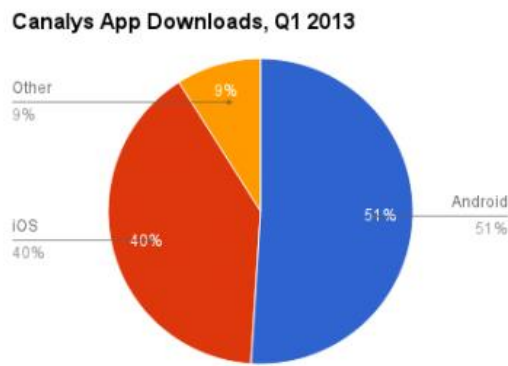

Ilustración 1. Cuota de mercado de aplicaciones (AMIPICI, 2014)

Entre 2010 y 2012 surgen distintos dispositivos celulares que ya incluyen soporte para la red WiMax, que a diferencia de la red WiFi que tiene alcance como máximo un kilómetro, la red WiMax puede llegar a alcances de hasta 80 kilómetros de cobertura de internet con velocidades de hasta 75 mbps.

Todos estos avances han puesto en literalmente al alcance de la mano cualquier tipo de información que esté disponible en la red, así como la posibilidad de llevar lo indispensable para poder comunicarnos y sustituir a la computadora con el teléfono celular. Esta condición ha llegado también al campo de la educación.

**La telefonía celular en la educación**

Hoy en día, muchos de los alumnos que estudian en línea a través de una computadora, acceden a plataformas educativas en la que son evaluados, por medio de un dispositivo de telefonía celular. La potencia de estos nuevos aparatos es cada vez más evidente, haciendo cada vez más posible que los teléfonos sean vistos como terminales de comunicación y trabajo, además de poder ejecutar un sinnúmero de aplicaciones que abarcan desde el trabajo hasta el ocio. Esto ha llevado a los estudiantes a explorar nuevas formas de aprendizaje, como lo que ahora se llama *m-learning* o *mobile learning*, el cual es el estudio y aprendizaje mediante dispositivos móviles, generalmente teléfonos celulares o tabletas inteligentes, generando el llamado *pad-learning*.

El uso de estos dispositivos ha hecho que la brecha digital existente con el uso de herramientas tecnológicas se haga cada vez más estrecha. En México, la Comisión Federal de Telecomunicaciones anunció en 2012 que la cifra de suscripciones de telefonía rebasaba los 100.6 millones, representando un 85.7 % de la población total (COFETEL, 2012). A 2014 y según datos de la Asociación Mexicana de Internet (AMIPCI, 2014), 5 de cada 10 internautas se conectan a través de su Smartphone (Ver Ilustración 2).

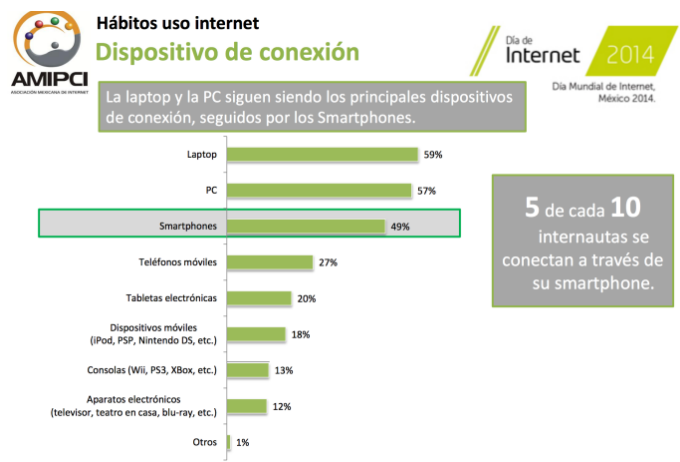

Ilustración 2. Uso de dispositivos para conectarse a Internet en México (AMIPCI, 2014)

El constante crecimiento de usuarios de tabletas y *smartphones*, nos lleva a que como tecnólogos educativos contemplemos seriamente la integración de estas herramientas en el quehacer educativo. Esto no necesariamente se convertirá al *m-learning* o *Tablet-learning*, o *pad-learning*. El *m-learning* es un camino al *u-learning* (Carbone, S.F.), término que comprende al aprendizaje en donde quiera que estemos (ubicuos) mediante cualquier dispositivo que pueda conectarse a internet.

Algunas de las universidades se han dado cuenta de este fenómeno, lo que les ha llevado a adaptar sus portales web para que puedan ser fácilmente accedidos por estos nuevos dispositivos, optimizando los recursos para que sea más fácil el proceso de aprendizaje por parte de los estudiantes.

Diversas instituciones educativas ya usan plataformas adecuadas para los dispositivos móviles, como por ejemplo la Universidad de Málaga, la Universidad Europea de Madrid y la Universidad Panamericana. En México existen algunas instituciones como la Universidad Autónoma de México, y el Instituto Tecnológico de Monterrey, por mencionar dos casos de ejemplo.

**Conclusiones**

Los teléfonos celulares con capacidades extendidas, los *smartphones*, son una solución efectiva ante la sustitución de los equipos de cómputo. Cada día es más común ver que los usuarios usan las características extras de los teléfonos celulares, tanto de hardware como en software.

Los diseñadores y desarrolladores de cursos deben contemplar que un importante número de internautas se conectan a la red por medio de sus dispositivos móviles, por lo que muchos de los contenidos educativos deberán ser adecuados a tecnologías escalables como el HTML 5 y optimizados para internet 2.0, desde la plataforma educativa correspondiente.

No podemos dar la espalda a nuevas tecnologías, y tampoco cerrar las posibilidades para que los estudiantes aprendan desde donde estén, teniendo dispositivos que les permiten navegar en internet. El reto como tecnólogos educativos se centrará en el desarrollo de estos contenidos y que sean compatibles en el futuro.

## REFERENCIAS